\documentclass[twocolumn,showpacs,aps]{revtex4}

\usepackage{color}   
\usepackage{graphicx}
\usepackage{dcolumn}
\usepackage{bm}

\begin{document}

\def\bbox#1{\hbox{\boldmath${#1}$}}
\def\blambda{{\hbox{\boldmath $\lambda$}}}
\def\eeta{{\hbox{\boldmath $\eta$}}}
\def\bxi{{\hbox{\boldmath $\xi$}}}
\def\bzeta{{\hbox{\boldmath $\zeta$}}}

\vspace*{-0.6cm}
\title{ Parton Momentum Distribution at the Moment of Jet-Parton
Collisions }

\author{Cheuk-Yin Wong\footnote{Email: wongc@ornl.gov}}

\affiliation{Physics Division, Oak Ridge National Laboratory, 
Oak Ridge, TN 37831, U.S.A.}

\date{\today}

\begin{abstract}

We extract the early parton momentum distribution using the STAR
Collaboration data of ridge particles associated with a near-side jet
in central AuAu collisions at $\sqrt{s_{NN}}=200$ GeV.  The ridge
particles are identified as medium partons kicked by the jet near the
surface and they carry direct information on the parton momentum
distribution at the moment of jet-parton collisions.  The extracted
parton momentum distribution has a thermal-like transverse momentum
distribution but a rapidity plateau structure with a relatively flat
rapidity distribution at mid-rapidities with sharp kinematic
boundaries at large rapidities that depend on the transverse momentum.

\end{abstract}

\pacs{ 25.75.Gz 25.75.Dw }

\maketitle


In central high-energy heavy-ion collisions, the state of the parton
medium during the early stage of a nucleus-nucleus collision is an
important physical quantity.  On the one hand, it furnishes
information for the investigation of the mechanism of parton
production in the early stages of the collision of two heavy nuclei.
On the other hand, it provides the initial information for the
evolution of the system toward the state of quark-gluon plasma.  Not
much is know about the early state of the partons from direct
experimental measurements.

Recently, the STAR Collaboration observed a $\Delta \phi$-$\Delta
\eta$ correlation of particles associated with a near-side jet in
central AuAu collisions at $\sqrt{s_{NN}}=200$ GeV at RHIC, where
$\Delta \phi$ and $\Delta \eta$ are the azimuthal angle and
pseudorapidity differences relative to a high-$p_t$ trigger particle
\cite{Ada05,Ada06,Put07,Wan07,Bie07}. The near-side correlations can
be decomposed into a ``jet'' component as fragmentation and radiation
products of the near-side jet at $(\Delta \phi, \Delta \eta)$$
\sim$(0,0), and a ``ridge'' component at $\Delta\phi$$ \sim$0 with a
ridge structure in $\Delta \eta$.

While many theoretical models have been proposed
\cite{Hwa03,Hwa07,Rom07,Vol05,Arm04,Won07,Won08,Won08a,Shu07,Pan07,Dum07,Gav08,Dum08a},
a momentum kick model was put forth to describe the ridge phenomenon
\cite{Won07,Won08,Won08a}.  The model assumes that a near-side jet
occurs near the surface and it kicks medium partons, loses energy
along its way, and fragments into the trigger particle and other
fragmentation and radiation products in the ``jet'' component.  The
kicked medium partons, each of which acquires a momentum kick from the
near-side jet, materialize by parton-hadron duality as ridge
particles.  They carry direct information on the parton momentum
distribution at the moment of jet-parton collisions.

We shall extract this parton momentum distribution for central AuAu
collisions at $\sqrt{s_{NN}}=200$ GeV using the STAR data
\cite{Ada05,Put07,Wan07}.  In a central AuAu collision, jet partons
with a distribution $dN_{ j}/d{\bf p}_j$ occur near the medium
surface.  The number of trigger particles reaching the near-side
detector with momentum ${\bf p}_{\rm trig}$ is
\begin{eqnarray}
\label{trig}
N_{\rm trig} \!\!= \!\!\int \! d{\bf p}_j \frac {dN_j}{d{\bf p}_j} 
\!\sum_{N\!=\!0}^{N_{\rm max}}
\!P_N(N) 
e^{-\zeta N} 
 D({\bf p}_{\rm trig};{\bf p}_j \!-\! \sum_{n = 1}^N {\bf q}_n),
\end{eqnarray}
where $N$ is the number of medium partons kicked by the jet, with a
maximum $N_{\rm max}$, and $P_N(N)$ is a geometry-dependent
probability distribution of $N$ normalized by $\sum_{N=0}^{N_{\rm
max}}P_N(N)=1$.  The factor $e^{-\zeta}$ describes the inelastic
attenuation of the jet for each jet-(medium parton) collision, and
$D({\bf p}_{a}; {\bf p}_b)$ is the fragmentation function for
fragmenting a ${\bf p}_a$ hadron from a ${\bf p}_b$ parton.  The
distribution of the associated kicked partons is therefore
\begin{eqnarray}
\label{rid}
\frac{dN_{\rm ridge}^{AA}}{d{\bf p}}
&=&\int d{\bf p}_j 
\frac{dN_j}{d{\bf p}_j} \sum_{N=1}^{N_{\rm max}}
P_N(N) e^{-\zeta N} 
\nonumber\\
&\times&
D({\bf p}_{\rm trig};{\bf p}_j - \sum_{n=1}^N {\bf q}_n)
\frac{2}{3} \sum_{n = 1}^N  \frac{dF_n}{d{\bf p}}({\bf q}_n), 
\end{eqnarray}
where $dF_n/d{\bf p}$ is the momentum distribution of the $n$-th
kicked medium parton, normalized to $\int d{\bf p} dF_n/d{\bf p} =1$,
and the factor (2/3) arises since the ridge particle detector accepts
only charged particles.  The formulation is greatly simplified upon
representing $dF_n/d{\bf p}$ by the average $dF/d{\bf p}$ and taking
the different momentum kicks ${\bf q}_n$ to be the average ${\bf q}$.
We then obtain
\begin{eqnarray}
\label{pertrig}
\frac{1}{N_{\rm trig}} \frac{dN_{\rm ridge}^{AA}}{d{\bf p}}
=\frac{2}{3}\langle N \rangle \frac{dF}{d{\bf p}} ({\bf q})=\frac{2}{3}\langle
N \rangle \frac{dF}{d{\bf p}} ({\bf q}),
\end{eqnarray}
where $\langle N \rangle$ is the average number of kicked partons per
trigger particle as determined from Eqs.\ (\ref{trig}) and (\ref{rid}),
\begin{eqnarray}
{\langle N \rangle} 
&=& \frac{1}{N_{\rm trig}}
\int {d{\bf p}_j}
\frac{dN_j}{d{\bf p}_j} \sum_{N=0}^{N_{\rm max}}
NP_N(N) e^{-\zeta N} 
\nonumber\\
&\times& 
D({\bf p}_{\rm trig};{\bf p}_j - \sum_{n=1}^N {\bf q}_n).
\end{eqnarray}
On the other hand, the distribution of the ``jet'' component of
associated fragmentation and radiation products is given by
\begin{eqnarray}
\label{jet1}
\frac{dN_{\rm jet}^{AA}}{d{\bf p}}
&=&\int d{\bf p}_j \frac{dN_j}{d{\bf p}_j} \sum_{N=0}^{N_{\rm max}} 
P_N(N)  e^{-\zeta N} 
\nonumber\\
&\times& 
D_2({\bf
p}_{\rm trig},{\bf p};{\bf p}_j - \sum_{n=1}^{N} {\bf q}_n),
\end{eqnarray}
where $D_2({\bf p}_a, {\bf p}_b ; {\bf p}_c)$ is the double
fragmentation function for fragmenting ${\bf p}_a$ and ${\bf p}_b$
hadrons from a jet parton of momentum ${\bf p}_c$.  Fragmentation
measurements \cite{Put07} suggest an approximate scaling relation
\begin{eqnarray}
D_2({\bf p}_{\rm trig}, {\bf p}; {\bf p}_c)
\approx D({\bf p}_{\rm trig}; {\bf p}_{c}) 
D_z({\bf p};{\bf p}_{\rm trig}),
\end{eqnarray}
where $D_z({\bf p} ; {\bf p}_{\rm trig} )$ is approximately the same
(within a factor of about 0.6 to 1.2) for dAu and AuAu collisions in
$2.5 <p_{t,\rm trig}< 6$ GeV (Fig. 5b and 5c of \cite{Put07}).
Applying this approximate scaling relation to Eq. (\ref{jet1}), the
jet component in an AA central collision per trigger is
\begin{eqnarray}
\label{jet}
\frac{1}{N_{\rm trig}} \frac {dN_{\rm jet}^{AA}}{d{\bf p}} \approx
D_z({\bf p};{\rm p}_{\rm trig}) \approx \frac{dN_{\rm jet }^{pp}}{d{\bf
p}}.
\end{eqnarray}
Because of the approximate nature of the above relation, we need to
make a quantitative check. In the region where the jet component has a
prominent appearance, as in Fig.\ 1(d) for $2< p_t < 4 $ GeV, the
total $dN_{\rm ch}/N_{\rm trig}d\Delta \eta$ distribution at $\Delta\eta\sim 0$
has indeed a similar shape as, but a peak magnitude about equal to,
the $pp$ near-side jet distribution.  Because the total yield is the
sum of the jet component and ridge yields and the ridge yield at
$\Delta\eta\sim 0$ is non-zero, the near-side jet component in AuAu
central collisions per trigger is thus an attenuated $pp$ near-side
jet distribution, as expected for production in an interacting
medium. If one assumes that fragmentation products lying deeper than
an absorption length from the surface are all absorbed, then the
average jet fragment transmission factor is $f_J=\int_0^\lambda
e^{-x/\lambda}dx/\lambda = 0.632$, which also leads to a reasonable
semi-empirical description of the experimental data as indicated below
in Figs. 1 and 2.  A ridge particle transmission factor $f_R$ can be
similarly introduced, but present measurements furnish information
only on the product $f_R\langle N \rangle$.  The sum of the
distributions (\ref{pertrig}) and (\ref{jet}), relative to the trigger
particle angles, is therefore given more precisely as
\begin{eqnarray}
\label{obs}
\left [ 
\frac{1}{N_{\rm trig}}
\frac{dN_{\rm ch} } 
{p_{t} dp_{t} d\Delta \eta  d\Delta \phi } \right ]_{\rm total}^{\rm AA} 
& & \!\!\!\!\!\!\!\!\!\!\! 
= \left [ f_R\frac{2}{3} \langle N \rangle \frac { dF } {p_t dp_t\,
d\Delta \eta\, d\Delta \phi} \right ]_{\rm ridge}^{\rm AA} \nonumber\\ 
& &  \!\!\!\!\!\!\!\!\!\!+
\left [ f_J  \frac { dN_{\rm jet}^{pp}} {p_t dp_t\, d\Delta \eta\, d\Delta
\phi} \right ]_{\rm jet}^{\rm AA} \!\!\!\!.
\end{eqnarray}
In the momentum kick model, the normalized ridge particle momentum
distribution $dF/d{\bf p}$ at ${\bf p}$ is related to the normalized
initial parton momentum distribution at a shifted momentum, ${\bf
p}_i={\bf p}-{\bf q}$, and we have \cite{Won07}
\begin{eqnarray}
\label{final}
\frac{dF}{ p_{t}dp_{t}d\eta d\phi} &=&\left [ \frac{dF}{
p_{ti}dp_{ti} dy_i d\phi_i } \frac{E}{E_i} \right ]_{{\bf p}_i={\bf p}-{\bf q}} \nonumber\\ &\times&
\sqrt{1-\frac{m^2}{(m^2+p_t^2) \cosh^2 y}},
\end{eqnarray}
where the factor $E/E_i$ insures conservation of particle number and
the last factor changes the rapidity to pseudorapidity distribution
\cite{Won94}. Therefore, the ridge particle momentum distribution $dF/
p_t dp_t d\Delta \eta d\Delta \phi$, relative to the trigger particle
angles, depends on ${\bf q}$ and the initial parton momentum
distribution.  The momentum kick ${\bf q}$ is expected to lie within a
narrow cone in the trigger particle direction.  To minimize the number
of parameters, we approximate ${\bf q}$ to lie along the trigger
particle direction.

The experimental $pp$ near-side jet data, shown as open circles in
Figs.\ 1 and 2, can be described well by the dash-dot curves in these
figures obtained from
\begin{eqnarray}
\label{jetfun}
\frac { dN_{\rm jet}^{pp}} {p_t dp_t\, d\Delta \eta\, d\Delta \phi}
\!\!&=& N_{\rm jet}
\frac{\exp\{(m-\sqrt{m^2+p_t^2})/T_{\rm jet}\}} {T_{\rm jet}(m+T_{\rm jet})}
\nonumber\\
&\times&
\frac{1}{2\pi\sigma_\phi^2}
e^{- {[(\Delta \phi)^2+(\Delta \eta)^2]}/{2\sigma_\phi^2} },
\end{eqnarray}
where 
$\sigma_\phi$=$\sigma_{\phi 0}\,{m_a}/{\sqrt{m_a^2+p_t^2}}$,
$\sigma_{\phi 0}$=0.5, $m_a$=1.1 GeV,
$N_{\rm jet}$=0.75, $m$=$m_\pi$, and $T_{\rm jet}$=0.55 GeV.
\begin{figure} [h]
\includegraphics[angle=0,scale=0.40]{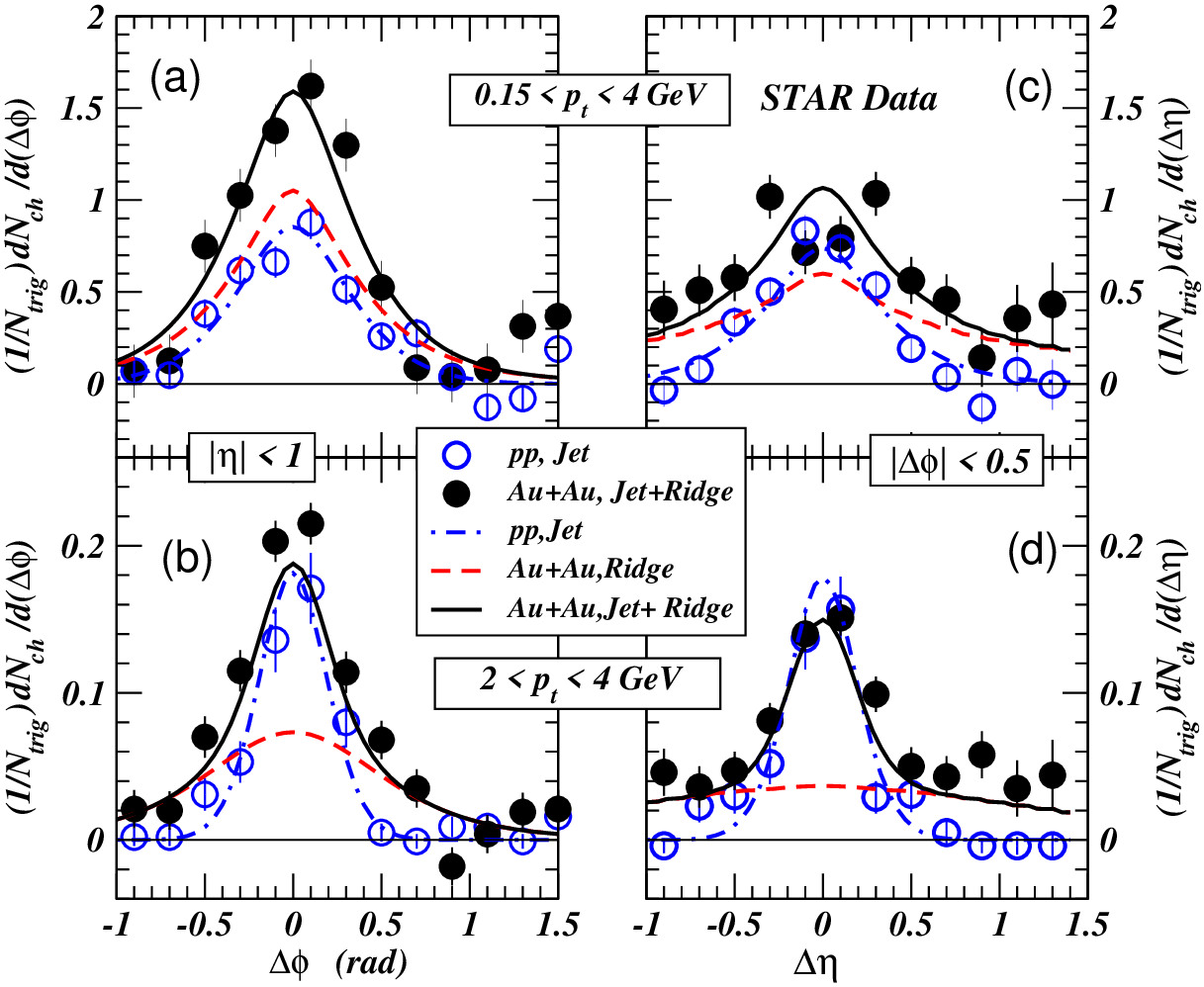}
\vspace*{0.0cm} 
\caption{ The symbols represent experimental data \cite{Ada05} and the
 curves theoretical results, for $pp$ and central AuAu collisions.
 (a) and (b) give the $dN_{\rm ch}/N_{\rm trig}d\Delta \phi$
 distributions.  (c) and (d) give the $dN_{\rm ch}/N_{\rm trig}d\Delta
 \eta$ distributions.  }
\end{figure}

The initial momentum distribution $dF/d{\bf p}_i$ of the medium
partons in Eq.\ (\ref{final}) is not yet a quantity that can be
obtained from first principles of QCD or the end-point, last-stage
bulk data.  It can however be extracted directly from the near-side
ridge data.  It was parametrized previously as $e^{-y_i^2/2\sigma_y^2}
\exp\{-\sqrt{m^2+p_{ti}^2}/T\}/\sqrt{m^2+p_{ti}^2}$, with $m$ taken to
be $m_\pi$ \cite{Won07}.  Although this is adequate for mid-rapidity
and high $p_t$ ridge particles \cite{Won07}, it leads to too large a
ridge yield both at $p_{t}\sim 1$ GeV (dotted curve in Fig. 2) and at
forward rapidities.  If the partons arise from a deconfined medium
with a finite transverse boundary, then the transverse parton momentum
distribution at small $p_t$ will be flattened from an exponential
distribution, as shown in Figs. 1 and 2 of \cite{Mos95}.  Transverse
distributions of this type can be described by replacing the
denominator $\sqrt{m^2+p_{ti}^2}$ with $\sqrt{m_d^2+p_{ti}^2}$ where
$m_d$ can be adjusted to lead to the correct ridge yield at $p_{t}\sim
1$ GeV.  The extracted transverse momentum distribution may provide
useful information to study the transverse radius of the deconfined
parton medium \cite{Mos95}.  The difficulty with the forward rapidity
region can be resolved by noting that the Gaussian rapidity
distribution of \cite{Won07} does not take into account the kinematic
boundary restrictions on phase space.  We can use a rapidity
distribution that retains the flatness at mid-rapidity but also
respects the kinematic boundaries at large rapidities and large
$p_t$. Accordingly, we parametrize the normalized initial parton
momentum distribution as
\begin{eqnarray}
\label{dis2}
\frac{dF}{ p_{ti}dp_{ti}dy_i d\phi_i}&=&
A_{\rm ridge} (1-x)^a 
\frac{ e^ { -\sqrt{m^2+p_{ti}^2}/T }} {\sqrt{m_d^2+p_{ti}^2}},
\end{eqnarray}
where $A_{\rm ridge}$ is a normalization constant defined by $\int
d{\bf p}_i dF/d{\bf p}_i =1$, $x$ is the light-cone variable
\cite{Won94}
\begin{eqnarray}
\label{xxx}
x=\frac{\sqrt{m^2+p_{ti}^2}}{m_b}e^{|y_i|-y_b},
\end{eqnarray}
$a$ is the fall-off parameter that specifies the rate of decrease of
the distribution as $x$ approaches unity, $y_b$ is the beam parton
rapidity, $m_b$ is the mass of the beam parton whose collision and
separation lead to the inside-outside cascade production of particles
\cite{Cas74,Won91,Won94}.  As $x \le 1$, there is a kinematic boundary
that is a function of $y_i$ and $p_{ti}$,
\begin{eqnarray}
\label{pty}
\sqrt{m^2+p_{ti}^2}=m_b e^{y_b-|y_i|}.
\end{eqnarray}
We expect $y_b$ to have a distribution centered around the nucleon
rapidity, $y_N=\cosh^{-1}(\sqrt{s_{NN}}/2m_N)$.  For lack of a
definitive determination, we shall set $y_b$ equal to $y_N$ and $m_b$
equal to $m$, pending their future experimental determination by
examining the ridge boundaries.

\begin{figure} [h]
\includegraphics[angle=0,scale=0.40]{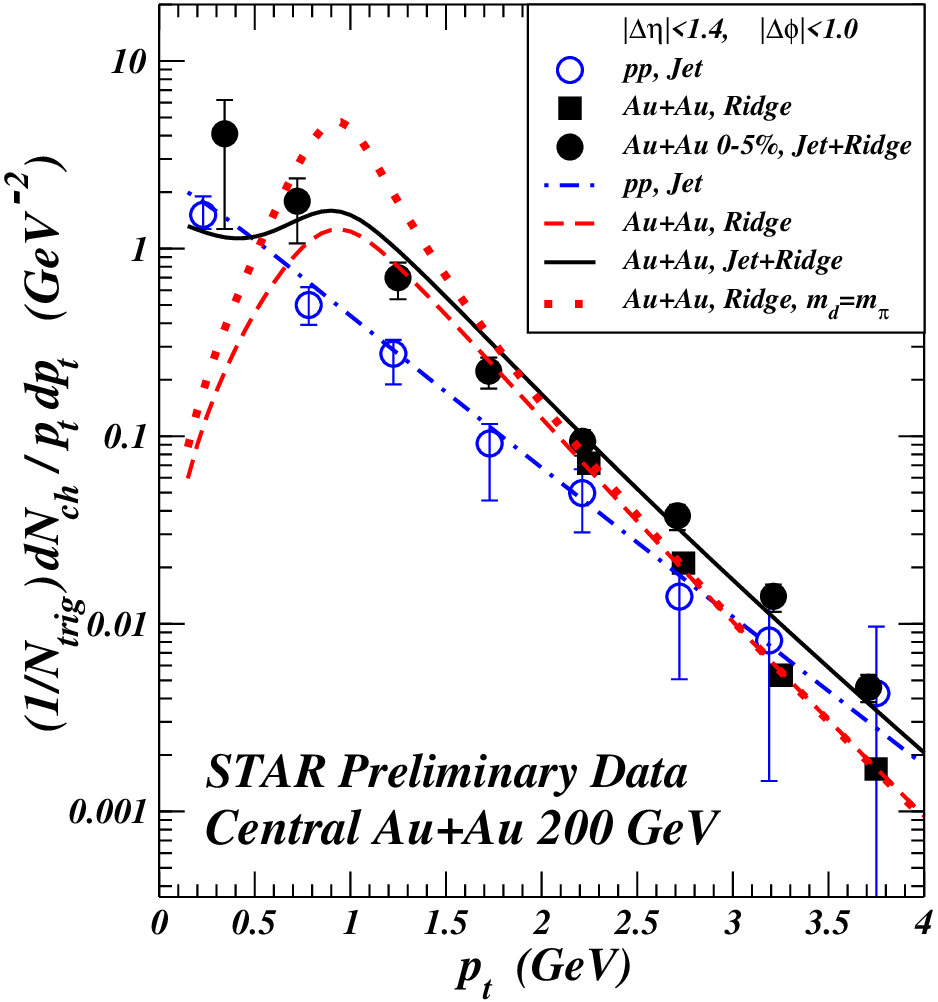}
\vspace*{0.0cm}
\caption{ The symbols represent STAR experimental data \cite{Ada05,Put07}
and the curves theoretical results of $dN_{\rm ch}/N_{\rm trig}p_t dp_t$, for
$pp$ and central AuAu collisions.}
\end{figure}

We find that the totality of the STAR near-side associated particle
data \cite{Ada05,Put07, Wan07}, in the region of $0.15 < p_t < 4$ GeV
and $0<|\eta|<3.9$ in central AuAu collisions at $\sqrt{s_{NN}}=200$
GeV, can be described by $|{\bf q}|=1.0$ GeV, $f_R \langle
N\rangle=4$, $a=$0.5, $T=$0.50 GeV, and  $m_d$=1 GeV.

We shall discuss the comparison of the experimental data with
theoretical results.  In Figs.\ 1, 2, and 3, the experimental total
associated particle yields \cite{Ada05,Wan07} are represented by solid
circles and the theoretical total and ridge yields by solid and dashed
curves, respectively.  In Fig.\ 1, comparison of theoretical and
experimental associated particle data indicates general agreement over
all azimuthal angles [Figs. 1(a) and 1(b)] and over all
pseudorapidities [Figs. 1(c) and 1(d)], for both $0.15 < p_t < 4 $ GeV
[Figs. 1(a) and 1(c)] and $2 < p_t < 4 $ GeV [Figs. 2(b) and 2(d)].
In Fig.\ 2, experimental ridge $dN_{\rm ch}/N_{\rm trig}p_tdp_t$ data
\cite{Put07} are shown as solid squares and they are calibrated by
using the data of Fig.\ 2 of \cite{Put07}.  Fig.\ 2 shows good
agreement between theoretical $dN_{\rm ch}/N_{\rm trig}p_t dp_t$ results with
experimental data.  Note that the theoretical ridge $dN_{\rm ch}/N_{\rm
trig}p_t dp_t$ has a peak at $p_t\sim |{\bf q}| \sim 1$ GeV, as a
result of the momentum kick.

\begin{figure} [h]
\includegraphics[angle=0,scale=0.40]{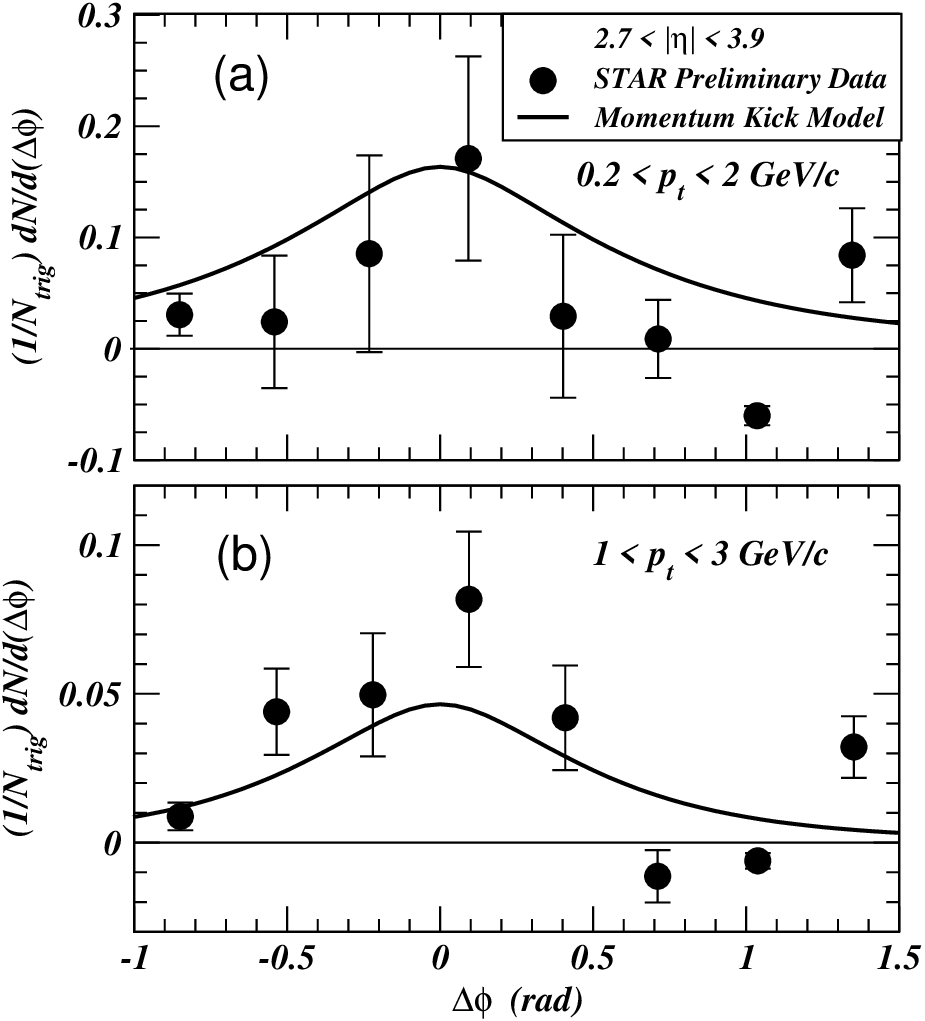}
\vspace*{0.0cm}
\caption{ STAR central AuAu collision azimuthal angle distribution
data at forward pseudorapidities \cite{Wan07} compared with results in
the momentum kick model (solid curves).  (a) is for $0.20 < p_t < 2$
GeV, and (b) is for $1 < p_t < 3$ GeV.  }
\end{figure}

We turn now to forward rapidities where preliminary experimental data
have been obtained for $2.7 <|\eta|< 3.9$ \cite{Wan07}. We note that
$dN_{\rm ch}/N_{\rm trig}d\Delta \phi d\Delta \eta $ at $\Delta \phi
\sim 0$ for $|\eta| < 1 $ in Fig.\ 1(a) is an order of magnitude
greater than the corresponding $dN_{\rm ch}/N_{\rm trig}d\Delta \phi
d\Delta \eta $ for 2.7$<$$|\eta|$$<$3.9 in Fig.\ 3(a). This implies a
substantial fall-off of ridge yield $dN_{\rm ch}/N_{\rm trig}\Delta
\phi d\Delta \eta$ at $\Delta \phi$$\sim$0 in going from
mid-rapidities to large rapidities.  Measurements at forward
rapidities in Fig. 3 contain events with large $\eta$ and $p_t$ that
are either already outside the kinematic limits or close to the
kinematic limits.  Therefore, even with the large errors, the forward
rapidity data in Fig. 3 are sensitive to the constraint of the
kinematic limits and the rate of fall-off of the initial parton
momentum distribution as specified by the fall-off parameter $a$.

\begin{figure} [h]
\includegraphics[angle=0,scale=0.40]{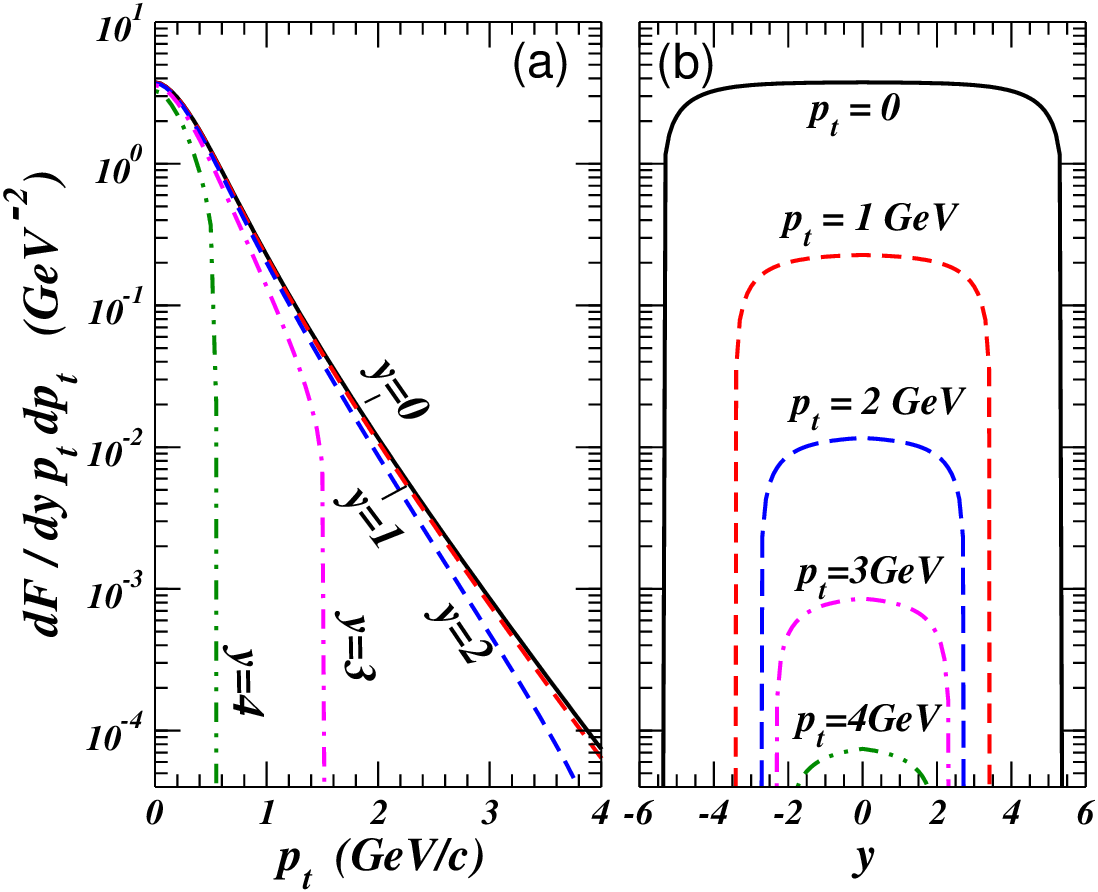}
\vspace*{0.0cm}
\caption{ Normalized initial parton momentum distribution $dF/dy p_t
dp_t$ extracted from the STAR Collaboration data
\cite{Ada05,Put07,Wan07}. (a) $dF/dy p_t dp_t$ as a function of $p_t$
for different $y$, and (b) $dF/dy p_t dp_t$ as a function of $y$ for
different $p_t$.  }
\end{figure}

The distribution (\ref{dis2}) with $a=$0.5, $T=$0.50 GeV, and $m_d$=1
GeV gives the extracted normalized initial parton momentum
distribution at the moment of jet-parton collision.  We show this
distribution $dF/p_t dp_t dy$ in Figs.\ 4(a) and 4(b), where $p_t$ and
$y$ are initial parton momentum variables.  In Fig. 4(b), the momentum
distribution as a function of $y$ for a fixed $p_t$ is essentially
flat near central rapidities and it extends to a maximum value of
$|y|_{\rm max}$ that depends on $p_t$.  The distribution decreases
rapidly as it approaches the kinematic limit. The rapidity plateau
structure suggests the inside-outside cascade picture of parton
production, as in the rapidity distribution of radiated particles when
a pair of color charges recede from each other
\cite{Cas74,Won91,Won94}.

In conclusion, the near-side ridge data in high-energy heavy-ion
collisions can be explained by the picture that a jet occurs near the
surface, kicks medium partons, loses energy, and fragments into the
trigger particle and other fragmentation products.  The kicked medium
partons materialize as ridge particles which carries information on
the early parton momentum distribution.  The extracted early parton
momentum distribution has a thermal-like transverse distribution but a
rapidity plateau structure whose width decreases as the transverse
momentum increases.  The early parton momentum distribution provides
valuable information for the mechanism of early parton production and
the later evolution of the system toward the state of quark-gluon
plasma.

After the manuscript was completed, predictions based on the present
momentum kick model for particle yields up to large $|\Delta \eta|$
associated with a near-side jet were found to agree well with
experimental measurements obtained by the PHOBOS Collaboration
\cite{Wen08}.

\vspace*{0.3cm} The author wishes to thank Profs. Fuqiang Wang,
J. Putsches, V. Cianciolo, Zhangbu Xu, Jiangyong Jia, and C. Nattrass
for helpful discussions and communications.  This research was
supported in part by the Division of Nuclear Physics, U.S. DOE, under
Contract No.  DE-AC05-00OR22725, managed by UT-Battle, LLC.

\end{document}